\documentclass[useAMs,usenatbib]{mn2e}  
\usepackage{graphicx} 
\usepackage{txfonts} 
\usepackage{natbib} 
\usepackage{hyperref} 
\usepackage{times} 
\def\gsim{\mathrel{\rlap{\lower4pt\hbox{\hskip1pt$\sim$}}
    \raise1pt\hbox{$>$}}}                
\hypersetup{ 
	bookmarks=true,         
    colorlinks=true,        
  linkcolor=blue,          
      citecolor=blue,        
      urlcolor=blue,           
}

\bibpunct{(}{)}{;}{a}{}{,}
\usepackage{aas_macros}

\begin{document} %

   \title[MWL Observations of 1ES 1011+496 in Spring 2008]{Multi-Wavelength Observations of the Blazar 1ES\,1011$+$496 in Spring 2008}

%
\author[M.~L.~Ahnen~et.~al.]{
\parbox{\textwidth}{
\small{
M.~L.~Ahnen$^{1}$,
S.~Ansoldi$^{2}$,
L.~A.~Antonelli$^{3}$,
P.~Antoranz$^{4}$,
A.~Babic$^{5}$,
B.~Banerjee$^{6}$,
P.~Bangale$^{7}$,
U.~Barres de Almeida$^{7}$\footnote{now at Centro Brasileiro de Pesquisas F\'isicas (CBPF\textbackslash{}MCTI), R. Dr. Xavier Sigaud, 150 - Urca, Rio de Janeiro - RJ, 22290-180, Brazil},
J.~A.~Barrio$^{8}$,
J.~Becerra Gonz\'alez$^{9}$\footnote{now at NASA Goddard Space Flight Center, Greenbelt, MD 20771, USA and Department of Physics and Department of Astronomy, University of Maryland, College Park, MD 20742, USA},
W.~Bednarek$^{10}$,
E.~Bernardini$^{11}$\footnote{Humboldt University of Berlin, Istitut f\"ur Physik  Newtonstr. 15, 12489 Berlin Germany},
B.~Biasuzzi$^{2}$,
A.~Biland$^{1}$,
O.~Blanch$^{12}$,
S.~Bonnefoy$^{8}$,
G.~Bonnoli$^{3}$,
F.~Borracci$^{7}$,
T.~Bretz$^{13}$\footnote{now at Ecole polytechnique f\'ed\'erale de Lausanne (EPFL), Lausanne, Switzerland},
E.~Carmona$^{14}$,
A.~Carosi$^{3}$,
A.~Chatterjee$^{6}$,
R.~Clavero$^{9}$,
P.~Colin$^{7}$,
E.~Colombo$^{9}$,
J.~L.~Contreras$^{8}$,
J.~Cortina$^{12}$,
S.~Covino$^{3}$,
P.~Da Vela$^{4}$,
F.~Dazzi$^{7}$,
A.~De Angelis$^{15}$,
G.~De Caneva$^{11}$,
B.~De Lotto$^{2}$,
E.~de O\~na Wilhelmi$^{16}$,
C.~Delgado Mendez$^{14}$,
F.~Di Pierro$^{3}$,
D.~Dominis Prester$^{5}$,
D.~Dorner$^{13}$,
M.~Doro$^{7}$\footnote{also at Universit\`a di Padova and INFN, I-35131 Padova, Italy},
S.~Einecke$^{17}$,
D.~Elsaesser$^{13}$,
A.~Fern\'andez-Barral$^{12}$,
D.~Fidalgo$^{8}$,
M.~V.~Fonseca$^{8}$,
L.~Font$^{18}$,
K.~Frantzen$^{17}$,
C.~Fruck$^{7}$,
D.~Galindo$^{19}$,
R.~J.~Garc\'ia L\'opez$^{9}$,
M.~Garczarczyk$^{11}$,
D.~Garrido Terrats$^{18}$,
M.~Gaug$^{18}$,
P.~Giammaria$^{3}$,
D.~Glawion (Eisenacher)$^{13}$,
N.~Godinovi\'c$^{5}$,
A.~Gonz\'alez Mu\~noz$^{12}$,
D.~Guberman$^{12}$,
Y.~Hanabata$^{20}$,
M.~Hayashida$^{20}$,
J.~Herrera$^{9}$,
J.~Hose$^{7}$,
D.~Hrupec$^{5}$,
G.~Hughes$^{1}$,
W.~Idec$^{10}$,
K.~Kodani$^{20}$,
Y.~Konno$^{20}$,
H.~Kubo$^{20}$,
J.~Kushida$^{20}$,
A.~La Barbera$^{3}$,
D.~Lelas$^{5}$,
E.~Lindfors$^{21}$,
S.~Lombardi$^{3}$,
F.~Longo$^{2}$,
M.~L\'opez$^{8}$,
R.~L\'opez-Coto$^{12}$,
A.~L\'opez-Oramas$^{12}$\footnote{now at Laboratoire AIM, Service d'Astrophysique, DSM\textbackslash{}IRFU, CEA\textbackslash{}Saclay FR-91191 Gif-sur-Yvette Cedex, France},
E.~Lorenz$^{7}$,
P.~Majumdar$^{6}$,
M.~Makariev$^{22}$,
K.~Mallot$^{11}$,
G.~Maneva$^{22}$,
M.~Manganaro$^{9}$,
K.~Mannheim$^{13}$,
L.~Maraschi$^{3}$,
B.~Marcote$^{19}$,
M.~Mariotti$^{15}$,
M.~Mart\'inez$^{12}$,
D.~Mazin$^{7}$,
U.~Menzel$^{7}$,
J.~M.~Miranda$^{4}$,
R.~Mirzoyan$^{7}$,
A.~Moralejo$^{12}$,
D.~Nakajima$^{20}$,
V.~Neustroev$^{21}$,
A.~Niedzwiecki$^{10}$,
M.~Nievas Rosillo$^{8}$,
K.~Nilsson$^{21}$\footnote{now at Finnish Centre for Astronomy with ESO (FINCA), Turku, Finland},
K.~Nishijima$^{20}$,
K.~Noda$^{7}$,
R.~Orito$^{20}$,
A.~Overkemping$^{17}$,
S.~Paiano$^{15}$,
J.~Palacio$^{12}$,
M.~Palatiello$^{2}$,
D.~Paneque$^{7}$,
R.~Paoletti$^{4}$,
J.~M.~Paredes$^{19}$,
X.~Paredes-Fortuny$^{19}$,
M.~Persic$^{2}$\footnote{also at INAF-Trieste},
J.~Poutanen$^{21}$,
P.~G.~Prada Moroni$^{23}$,
E.~Prandini$^{1}$\footnote{also at ISDC - Science Data Center for Astrophysics, 1290, Versoix (Geneva)},
I.~Puljak$^{5}$,
R.~Reinthal$^{21,*}$,
W.~Rhode$^{17}$,
M.~Rib\'o$^{19}$,
J.~Rico$^{12}$,
J.~Rodriguez Garcia$^{7}$,
S.~R\"ugamer$^{13,*}$,
T.~Saito$^{20}$,
K.~Satalecka$^{8}$,
V.~Scapin$^{8}$,
C.~Schultz$^{15}$,
T.~Schweizer$^{7}$,
S.~N.~Shore$^{23}$,
A.~Sillanp\"a\"a$^{21}$,
J.~Sitarek$^{10}$,
I.~Snidaric$^{5}$,
D.~Sobczynska$^{10}$,
A.~Stamerra$^{3}$,
T.~Steinbring$^{13}$,
M.~Strzys$^{7}$,
L.~Takalo$^{21}$,
H.~Takami$^{20}$,
F.~Tavecchio$^{3}$,
P.~Temnikov$^{22}$,
T.~Terzi\'c$^{5}$,
D.~Tescaro$^{9}$,
M.~Teshima$^{7}$,
J.~Thaele$^{17}$,
D.~F.~Torres$^{24}$,
T.~Toyama$^{7}$,
A.~Treves$^{25}$,
V.~Verguilov$^{22}$,
I.~Vovk$^{7}$,
J.~E.~Ward$^{12}$,
M.~Will$^{9}$,
M.~H.~Wu$^{16}$,
R.~Zanin$^{19}$
}
\textit{(The MAGIC Collaboration)}\\
F.~Lucarelli$^{27}$, 
C.~Pittori$^{27}$,
S.~Vercellone$^{28}$ \textit{for the AGILE team}\\
A.~Berdyugin$^{29}$,
M.~T.~Carini$^{30}$, 
A.~L\"ahteenm\"aki$^{31,32}$,
M.~Pasanen$^{29}$,
A.~Pease$^{30}$, 
J.~Sainio$^{29}$,
M.~Tornikoski$^{31}$,
R.~Walters$^{30}$ 
\vspace{0.25cm}\\
(Affiliations can be found at the end of the article).
}
}

\date{Received 2015; accepted 2016; Draft version \today}
\pagerange{\pageref{firstpage}--\pageref{lastpage}} \pubyear{2016}
\maketitle
\label{firstpage}
\begin{abstract}
   The BL Lac object 1ES\,1011$+$496 was discovered at Very High Energy (VHE, E$>$100GeV) $\gamma$-rays by MAGIC in spring 2007. Before that the source was little studied in different wavelengths. Therefore a multi-wavelength (MWL) campaign was organized in spring 2008. Along MAGIC, the MWL campaign included the Mets\"ahovi radio observatory, Bell and KVA optical telescopes and the \emph{Swift} and \emph{AGILE} satellites. MAGIC observations span from March to May, 2008 for a total of 27.9 hours, of which 19.4 hours remained after quality cuts. The light curve showed no significant variability yielding an integral flux above 200\,GeV of $(1.3\,\pm\,0.3)\,\times\,10^{-11}$ photons $\mathrm{cm}^{-2}\mathrm{s}^{-1}$. The differential VHE spectrum could be described with a power-law function with a spectral index of $3.3\pm0.4$. Both results were similar to those obtained during the discovery. \emph{Swift} XRT observations revealed an X-ray flare, characterized by a harder-when-brighter trend, as is typical for high synchrotron peak BL Lac objects (HBL). Strong optical variability was found during the campaign, but no conclusion on the connection between the optical and VHE $\gamma$-ray bands could be drawn. The contemporaneous SED shows a synchrotron dominated source, unlike concluded in previous work based on non-simultaneous data, and is well described by a standard one--zone synchrotron self--Compton model. We also performed a study on the source classification. While the optical and X-ray data taken during our campaign show typical characteristics of an HBL, we suggest, based on archival data, that 1ES\,1011$+$496 is actually a borderline case between intermediate and high synchrotron peak frequency BL Lac objects.
\end{abstract}



\section{Introduction}

Blazars are active galactic nuclei (AGNs) with relativistic jets oriented close to our line of sight.
The blazar spectral energy distribution (SED) is characterised by two broad peaks of which the lower energy one is believed to originate from synchrotron emission of electrons in the jet. The higher energy peak is most commonly explained by inverse-Compton scattering of either the synchrotron (synchrotron self Compton - SSC, see e.g. ~\citet{maraschi1992, costamante2002}) or external (external Compton - EC, ~\citep{dermer1993, ghisellini2005}) seed photons by the electrons and positrons in the jet. Hadronic models, where the $\gamma$-rays are produced directly by proton-synchrotron emission or via pion decay ~\citep{mannheim1993, mycke2003,2015A&A...573A...7W}, have also been suggested. BL Lac objects, a type of blazars with weak or no optical spectral lines, are subdivided into low, intermediate and high synchrotron peak BL Lac objects (LBL, IBL and HBL, respectively) according to the frequency of the first peak which in the case of HBLs is located in the UV to hard X-ray regime ~\citep[e.g.][]{padovani1995, sambruna1996}.  
 
Blazars show flux and spectral variability at all wavelengths on time scales ranging from a few minutes to several months \citep[e.g.][]{giommi1990,nieppola2007,albert2007a}. Therefore, in order to shed light on the VHE emission mechanisms and its origin in blazars, simultaneous observations of these sources at different flux states and across multiple wavelengths are required. It is particularly important to study correlations between flux and spectral variations in different energy bands.

The 1ES\,1011$+$496 MWL campaign discussed in this paper took place in spring 2008 before the launch of the \emph{Fermi} satellite, with MAGIC, AGILE, \emph{Swift} XRT and UVOT, KVA, Bell and Mets\"ahovi telescopes observing the source. While the broad-band SEDs of the source have been presented by several authors (e.g. ~\citet{albert2007, tavecchio2010, abdo2010, zhang2012} and~\citet{giommi2012}), these were not based on simultaneous data or did not include VHE $\gamma$-ray observations. Part of the results of this campaign have already been published in~\citet{reinthal2011,reinthal2012b}. Another similar campaign was conducted in 2008 concentrating on the HBL object 1ES\,2344$+$514 \citep{aleksic2013}.

1ES\,1011$+$496 is a BL Lac object, first detected as an X-ray source \citep{elvis1992}, located at a medium redshift of $z=$0.212 ~\citep{albert2007}. It was discovered at VHE by MAGIC in 2007 following an optical high state reported by the Tuorla Blazar Monitoring Program\footnote{\href{http://users.utu.fi/kani/1m}{http://users.utu.fi/kani/1m}} ~\citep{albert2007}. At the time of the discovery, 1ES\,1011$+$496 was the most distant source known to emit VHE $\gamma$-rays. Previous observations of this source at VHE showed only a hint of a signal (see e.g. ~\citet{albert2008a}) and the results presented here constitute the first follow-up observation of 1ES\,1011$+$496.

HBLs are the most numerous extragalactic VHE $\gamma$-ray sources. 
\citet{2011ApJ...742...27L} found using MOJAVE\footnote{www.physics.purdue.edu/MOJAVE/} 15\,GHz Very Long Baseline Interferometry (VLBI) data that these sources are distinguished from other blazar populations by lower-than-average radio core brightness temperatures and lack of high linear polarization in the core. \citet{2010ApJ...715..429A} found that the GeV spectra of the HBLs are essentially compatible with power laws.
1ES~1011+496 has multiple classifications in the literature. Multi-band monitoring between 2005 and 2010 by the McGraw-Hill Telescope
revealed a peak located in the optical regime ($\sim$2--3\,eV), indicating an IBL nature. A trend of the peak location shifting to higher energies with increasing flux was also identified ~\citep{bottcher2010}. Despite the long observing period, the authors note that the source has been observed mostly in moderately faint states. However, the object has historically been classified as an HBL object ~\citep{donato2001, nieppola2006, abdo2010}. We combine the different archival observations with data collected during our campaign in a new, consistent interpretation of the nature of the source.

The following section will be devoted to the instruments participating in the MWL campaign, their observations as well as their data analysis description. The results of these analyses are reported and compared to previous results in Section~3, and the MWL light curve, quasi-simultaneous SEDs and source classification are discussed in Section~4. A summary of the findings presented in this paper and concluding remarks are given in Section~5.

\section{Multi-wavelength observations and participating instruments}
The campaign was centered around common observation windows of the AGILE satellite and the MAGIC telescope in spring 2008. Additional MWL coverage was provided by the Mets\"ahovi radio telescope in the radio band, by the KVA and Bell telescopes in the optical waveband, and in the X-rays by the \emph{Swift} satellite. 
The 2008 MWL campaign was the first to incorporate VHE coverage for this source.

\subsection{MAGIC telescope}
MAGIC (Major Atmospheric Gamma-ray Imaging Cherenkov) is a system of two 17\,m Imaging Atmospheric Cherenkov Telescopes (IACTs) located on the Canary island of La Palma, Spain, at $\sim$2200\,m above sea level. At the time of the 2008 campaign the second telescope was still under construction and observations were performed using MAGIC-I only, having been in operation since 2004. Thanks largely to its 236\,m$^{2}$ reflective area, MAGIC-I achieved an energy threshold of $\sim$60\,GeV 
-- the lowest of any IACT at the time. It reached a sensitivity of $\sim$1.6 per cent of the Crab nebula flux (5\,$\sigma$ in 50\,h) $>$200\,GeV with an energy resolution of $\sim$20-30 per cent and an angular resolution of $\sim$0.1$^{\circ}$ ~\citep{albert2008}.

MAGIC was able to observe the object on 25 nights between 2008 March 4th and May 24th. The observation period was plagued by poor weather conditions at La Palma with frequent clouds, rainfall, strong wind and calima (dust from the Sahara) towards the end of the observation window. Nevertheless a total of 27.9 hours of data between zenith angles of 20$^\circ$ and 37$^\circ$ were collected, of which 8 hours had to be removed due to adverse weather conditions. The remaining 20 hours of data were not significantly affected by bad weather and survived the quality cuts.
The observations were conducted in wobble mode with the telescope alternating between two sky positions offset 0.4$^\circ$ from the source, allowing for simultaneous recording of ON and OFF data ~\citep{daum1997}.

The data were analysed using the MAGIC standard analysis package "MARS"~\citep{moralejo2009}. The images were cleaned using the timing information of the showers~\citep{aliu2009} and absolute cleaning levels of 6 photoelectrons (for the so-called "core pixels") and 3 photoelectrons (for "boundary pixels"). The cleaned images were then parametrised according to parameters described in~\citet{hillas1985}. $\gamma$-ray and background events were separated on basis of a Random Forest regression method~\citep{albert2008} and a cut in $\alpha$, the angle between the major shower axis and the line determined by the centre of gravity and the source position on the camera. Energy look up tables were used for the energy reconstruction. 
The results presented here have been confirmed internally by an independent analysis.

\subsection{AGILE space telescope}
AGILE (Astrorivelatore Gamma ad Immagini LEggero) is a scientific mission funded by the Italian Space Agency (ASI) dedicated to the observation of astrophysical sources of high energy astrophysics~\citep{tavani2009}. Launched on April 23, 2007 in a low-Earth orbit optimized for low particle background (with initial altitude of about 550 km), AGILE is working nominally after almost 8 years of operations.

In this paper, we have analysed data collected during the 2008 MWL campaing by the main AGILE instrument, the Gamma-Ray Imaging Detector (GRID).
The AGILE-GRID consists of a silicon-tungsten tracker, a cesium iodide mini-calorimeter and an anticoincidence system made of segmented plastic scintillators, and it is sensitive in the energy range from 30 MeV -- 50 GeV. The use of the silicon strip technology allows to have good performance for the $\gamma$--ray GRID imager, approximately a small cube of about 60 cm size, which achieves an effective area of the order of 500~cm$^2$ at several hundreds MeV, an angular resolution (at 68\% containment radius) of about 4.3$^\circ$ at 100 MeV, decreasing below 1$^\circ$ for energies above 1 GeV~\citep{2013A&A...558A..37C}, a
large field of view 
 of about $\sim 2.5$ sr, as well as accurate timing, positional and attitude information (source location accuracy 5 -- 10 arcmin for intense sources with S/N $ \gsim 10$).

During its first period of data taking (about two years), the AGILE satellite was operated in “pointing observing mode”, and the corresponding AGILE data are grouped in Observation Blocks. 
The time period covered by the 2008 MWL campaign includes the AGILE Observation Blocks 5500, 5510, 5520 and 5530, publicly available from the AGILE Data Center (ADC) web pages\footnote{http://agile.asdc.asi.it}. The source 1ES 1011+496 was observed, on average, at about 40$^\circ$ off-axis from the mean AGILE pointing direction in the two time windows: March 30 -- April 10, 2008 and April 30 -- May 10, 2008.

AGILE data were analysed using the latest scientific software (AGILE SW 5.0 SourceCode) and in-flight calibrations (I0023) publicly available at the ADC site. Counts, exposure and diffuse $\gamma$-ray background 
maps were created for energies $ E > 100$ MeV including $\gamma$-ray events collected up to 60$^\circ$ off-axis. Events collected during passages over the South Atlantic Anomaly, and regions within 10$^\circ$ from the Earth limb were rejected. 
In order to derive the source flux (or flux upper limits) in the full AGILE-GRID energy band (100 MeV -- 50 GeV), we ran the AGILE point source analysis software based on the Maximum Likelihood 
technique with a radius of analysis of 10$^\circ$.

\subsection{Swift XRT and UVOT}
\emph{Swift} is a MWL observatory 
launched into Low Earth Orbit in November 2004~\citep{gehrels2004}. The satellite is equipped with three telescopes: the Burst Alert Telescope (BAT,~\citet{barthelmy2005}) covering the 15\,--\,150\,keV energy range, the X-ray Telescope (XRT, ~\citet{burrows2005}) operating in the 0.2\,--\,10\,keV energy band and the UV/Optical Telescope (UVOT, ~\citet{roming2005}) for simultaneous UV and optical observations between 180 and 600\,nm.

\emph{Swift} XRT observed the source for 10 days between April 28 and May 8, 2008 (the results are summarized in Table~\ref{XRTpl}). The usable exposure times ranged from $\sim$200\,s to 2\,ks, while the shortest exposure was insufficient for deriving an X-ray flux and was discarded. The XRT data were processed with standard procedures using the FTOOLS task XRTPIPELINE (version 0.12.8) distributed by HEASARC within the HEASOFT package (v.6.15). Events with grades 0\,--\,12 were selected \citep[see][]{burrows2005} and response matrices of \emph{Swift} CALDB release 20071106 were used.

XRT observations were taken in photon-counting mode (PC) and are affected by a moderate pile-up due to the source having been brighter than expected. It was evaluated following the standard procedure\footnote{http://www.swift.ac.uk/analysis/xrt/pileup.php}, resulting in a piled-up region with a radius of $\sim$\,7\,arcsec. This region was masked extracting the signal within an annulus with inner radius of 3 pixels (7.1\,arcsec) and outer radius of 25 pixels (59\,arcsec). 

The spectra were extracted from the corresponding event files and binned using GRPPHA to ensure a minimum of 25 counts per energy bin, in order to guarantee reliable $\chi^2$ statistics ~\citep{gehrels1986}. Spectral analyses were performed using XSPEC version 12.8.1. The spectral index was determined using an absorbed power-law fit ($f_0 \times E^{-\Gamma} \times \mathrm{e}^{-\tau}$) from 0.3\,--\,10\,keV, with the optical depth $\tau$ being the product of the hydrogen column density $N_\mathrm{H}$ and the energy-dependent photoelectric cross section $\sigma\left(E\right)$. $N_\mathrm{H}$ was fixed to the Galactic value in the direction of the source of $8.4 \times 10^{19}\,\mathrm{cm}^{-2}$ evaluated from the Leiden/Argentine/Bonn (LAB) survey of galactic HI ~\citep{kalberla2005}. Since some daily data sets showed hints of spectral curvature, also fits using a log-parabola model ($f_0 \times E^{-\left(\Gamma+\beta \log_{10}\left(E\right)\right)} \times \mathrm{e}^{-\tau}$) were performed. However, for the majority of the cases the log-parabola fit was not significantly preferred by a logarithmic likelihood ratio test over the simple power-law model (see Table~\ref{XRTpl}). Therefore, the simple power-law results were used as a common basis.

\begin{table*}
 \centering
 \caption{\emph{Swift} XRT flux and spectral results. $\Gamma$ and $\chi_\mathrm{red}^2$/d.o.f. are the spectral index and reduced $\chi^2$ over the number of degrees of freedom of the simple power-law fit. $L$ denotes the likelihood ratio of this power-law fit when compared to a log-parabolic fit.}
  \begin{tabular}{@{}llccccc@{}}
  \hline
  \hline
Obs.\ ID & MJD start & Exposure & Flux 0.5-10\,keV & $\Gamma$ & $\chi_\mathrm{red}^2$/d.o.f.\ & $L$ \\
 & & [ks] & [$10^{-11}\,\mathrm{ph}\,\mathrm{cm}^{-2}\,\mathrm{s}^{-1}$] & & & [\%] \\
  \hline
	35012008 & 54584.8868 & 2.00 & $4.73 \pm 0.18$ & $2.28 \pm 0.06$ & 1.06/55 & 98.0 \\
	35012009 & 54586.8250 & 0.19 & $6.7^{+1.3}_{-0.8}$ & $2.00 \pm 0.30$ & 1.2/3\phantom{0} & 53.5 \\
	35012011 & 54588.9118 & 0.86 & $6.22 \pm 0.32$ & $2.19 \pm 0.08$ & 0.95/31 & 97.9 \\
	35012012 & 54589.8993 & 1.51 & $5.47 \pm 0.21$ & $2.27 \pm 0.07$ & 0.94/45 & 74.4 \\
	35012014 & 54590.8972 & 1.67 & $5.33 \pm 0.21$ & $2.30 \pm 0.07$ & 0.81/42 & 74.2 \\
	35012013 & 54591.9007 & 1.37 & $4.67 \pm 0.19$ & $2.37 \pm 0.06$ & 1.81/48 & 98.9 \\
	35012015 & 54592.9049 & 1.79 & $4.53 \pm 0.17$ & $2.47 \pm 0.06$ & 1.34/51 & 98.9 \\
	35012016 & 54593.0382 & 1.87 & $4.62 \pm 0.14$ & $2.41 \pm 0.06$ & 1.22/54 & 98.5 \\
	35012017 & 54594.3306 & 1.77 & $3.73 \pm 0.15$ & $2.44 \pm 0.07$ & 1.19/42 & \ldots \\

  \hline
  \end{tabular}
 \label{XRTpl}
\end{table*}

\emph{Swift} UVOT cycled  through each of the optical and the UV pass bands \textit{V, B, U, UVW1, UVM2, UVW2}. The source counts were extracted from a circular region 6 arcsec-sized centred on the source position, while the background was extracted from a larger circular nearby source-free region. These data were processed with the {\tt uvotmaghist} task of the HEASOFT package. The observed magnitudes have been corrected for Galactic extinction $E(B-V) = 0.012$\,mag \citep{schlafly2011} using the extinction curve from \citet{fitzpatrick1999}. 
The host-galaxy flux contributes to the observed brightness in the \textit{V}- and \textit{B}-bands, however no values for the contribution were found in the literature. Therefore, the contributions were estimated from the \textit{R}-band value from ~\citet{nilsson2007} (the host galaxy contribution in the \textit{R}-band is $F_R$ =0.49\,mJy within an aperture of 7.5\,arcsec measured with a seeing of 3.0\,arcsec) using the galaxy colours at $z = 0.2$ from ~\citet{fukugita1995} resulting in $F_V$ = 0.27\,mJy and $F_B$ = 0.07\,mJy.

The magnitudes measured in the UV filters were converted to units of $\mathrm{erg}\,\mathrm{cm}^{-2}\,\mathrm{s}^{-1}$ using the photometric zero points as given in ~\citet{breeveld2011} and effective wavelengths and count-rate-to-flux ratios of GRBs from the \emph{Swift} UVOT CALDB 02 (v.20101130). ~\citet{raiteri2010} noted that these ratios are not necessarily applicable to BL Lac objects, due to their different spectrum and a \textit{B}\,--\,\textit{V} value typically larger than the applicable range and calculated a new calibration. 
Following the argumentation in ~\citet{aleksic2013} we did not apply this new calibration, but increased the error of the \textit{UVW2} count-rate-to-flux ratio from $\sim$\,2.2 to 13 per cent to account for a potential change in this value as large as found by ~\citet{raiteri2010}. However the actual uncertainty should be much below that, considering that some (if not most) of the difference between the ratios arises solely from using new effective wavelengths, which were not used in our work.

\subsection{KVA telescope and the Tuorla Blazar Monitoring Program}
\begin{figure}
 \vspace{5mm}
 \centering
 \includegraphics[width=.7\linewidth, angle=270]{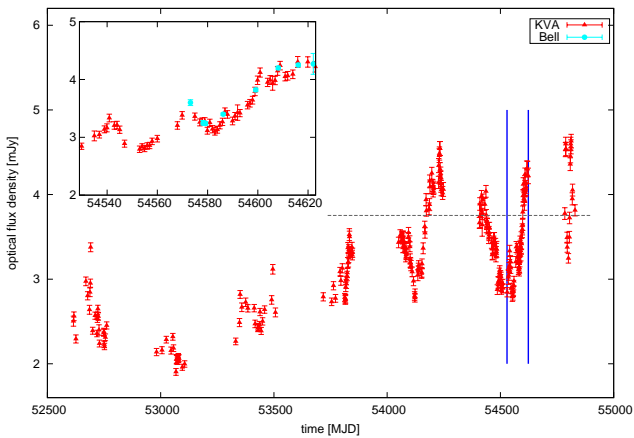}
 \caption{Long-term optical light curve of 1ES\,1011$+$496 until the end of 2008 from the Tuorla Blazar Monitoring Program (red dots). The fluxes are measured in the \textit{R}-band and they have not been host-galaxy subtracted. The horizontal dashed grey line represents the flare limit flux from ~\citet{reinthal2012a} to which the host galaxy contribution has been added to allow for direct comparison to the data points. The vertical blue lines denote the beginning and the end of the MWL campaign. The inset shows a zoom into the MWL campaign from the beginning of the MAGIC observations on March 4 until the last Bell observation on June 5. The Bell data are represented by cyan filled squares. Both data are binned daily.}
 \label{1011_LC_long}
\end{figure}

The bulk of the optical data presented in this paper were provided by the Tuorla Blazar Monitoring Program, which is operated as a support program to the MAGIC observations. The program was started at the end of 2002 and uses the Tuorla 1\,m telescope (located in Tuorla, Finland) and the Kungliga Vetenskapsakademien (KVA) telescope (located at the Roque de los Muchachos observatory on La Palma) to monitor candidate (from ~\citet{costamante2002}) and known TeV blazars in the optical waveband. The purpose of this monitoring is to study the long-term optical behaviour of the selected sources and provide alerts to MAGIC on high states of these objects in order to trigger follow-up VHE observations. 1ES\,1011$+$496 has been observed since the beginning of the program.

The KVA telescope consists of two tubes. The larger of the two, a 60\,cm reflector, is equipped with a CCD polarimeter capable of polarimetric measurements in \textit{BVRI} bands using a plane-parallel calcite plate and a super-achromatic $\lambda/$2 retarder ~\citep{piirola}, while the smaller one, a 35\,cm Celestron Schmidt-Cassegrain, is used for photometry. They are operated remotely from Finland. The photometric measurements are usually done in the \textit{R}-band. For this campaign the \textit{B}- and \textit{V}-band observations were also performed. During the campaign we also performed polarimetric observations on three nights without a filter to improve the signal-to-noise ratio of the observations. 

The observations have been conducted using differential photometry, i.e.\ by having the target and the calibrated comparison stars on the same CCD images ~\citep{fiorucci1996}. The magnitudes of the source and comparison stars are measured using aperture photometry adopting an aperture radius of 7.5\,arcsec and converted to linear flux densities according to the formula $F\mathrm = F_{0}\, \times\, 10^{-0.4\mathrm{mag}}$ Jy, where $F_{0}$ is a filter-dependent zero point ($F_{0}$ = 3080\,Jy in the \textit{R}-band, $F_{0}$ = 3640\,Jy in the \textit{V}-band and $F_{0}$ = 4260\,Jy in the \textit{B}-band, from ~\citet{bessell1979}). 

The polarimetric data are analysed using the standard procedures with a semiautomatic software specially developed for polarization monitoring purpose. In short, the normalized Stokes parameters and the degree of polarization and position angle were calculated from the intensity ratios of the ordinary and extraordinary beams using the standard formula (e.g. \citet{2007ASPC..364..495L}). 

In order to obtain the AGN core flux, emission from the host galaxy and possible nearby stars that contribute to the overall flux have to be subtracted from the measured value. ~\citet{nilsson2007} determined these contributions in the optical \textit{R}-band for many of the sources in the Tuorla monitoring list and in the case of 1ES\,1011$+$496 a host galaxy flux of (0.49 $\pm$ 0.02)\,mJy has to be subtracted from the measured \textit{R}-band flux. As discussed in the previous section, host galaxy contributions in the \textit{V}- and \textit{B}-bands had to be estimated from the \textit{R}-band value and amounted to \textit{$F_V$} = 0.27\,mJy and \textit{$F_B$} = 0.07\,mJy, respectively. Also these observed magnitudes have been corrected for Galactic extinction \textit{E(B-V)} = 0.012\,mag \citep{schlafly2011} using the extinction curve from \citet{fitzpatrick1999}.

At the time of the campaign there were no published \textit{R}-, \textit{B}- and \textit{V}-band magnitudes of the five comparison stars for the field of 1ES\,1011$+$496. We therefore calibrated the magnitudes ourselves using the comparison stars of S5\,0716$+$714 with known magnitudes observed in the same photometric nights. The results are given in Table~\ref{standards}, the numbering of the stars follows that of 
~\citet{bottcher2010}. The derived magnitudes are in good agreement with those in ~\citet{bottcher2010}, deviating typically by less than 2\,$\sigma$. 

\begin{table}
\centering
\caption{\small{Calibrated magnitudes of comparison stars.}}
\begin{tabular}{cccc}
\hline
\hline
Star & B & V & R \\
\hline
1 & $14.68 \pm 0.05$ & $13.87 \pm 0.04$ & $13.40 \pm 0.02$ \\
2 & $15.00 \pm 0.05$ & $14.43 \pm 0.04$ & $14.04 \pm 0.02$ \\
3 & $16.63 \pm 0.05$ & $15.88 \pm 0.04$ & $15.44 \pm 0.02$ \\
4 & $14.62 \pm 0.05$ & $14.32 \pm 0.04$ & $14.01 \pm 0.02$ \\
5 & $16.30 \pm 0.05$ & $15.73 \pm 0.04$ & $15.42 \pm 0.02$ \\
\hline
\end{tabular}
\label{standards}
\end{table}

The source has shown strong variability ever since the beginning of the monitoring, as can be seen from the long-term optical light curve in Fig.~\ref{1011_LC_long}.

\subsection{Bell telescope}
Observations from the Western Kentucky University Bell Observatory were obtained with a 60\,cm telescope and an Apogee AP6ep CCD camera, through an \textit{R}-band filter. The source was observed on 8 nights between April 17 and June 5, 2008. Differential photometry was performed between the blazar and published comparison stars on the same CCD frame. The comparison stars and apertures used were the same as for KVA to ensure comparability between the two instruments.

\subsection{Mets\"ahovi radio telescope}
The 37\,GHz observations were made with the Mets\"ahovi radio telescope located in Kylm\"al\"a, Finland. The telescope has a 13.7\,m diameter ESSCO design antenna placed inside a radome. The measurements were made with a 1\,GHz-band dual beam receiver centred at 36.8\,GHz with the antenna half power beam width of 2.4\,arcmin and a beam separation of 6.0\,arcmin. The telescope detection limit at 37\,GHz is $\sim$0.2\,Jy under optimal conditions. For a more comprehensive overview of the telescope, the observation methods and the data analysis procedure, refer to e.g.\ ~\citet{terasranta1998}.

Mets\"ahovi measurements were hindered by bad weather during the campaign. The source was observed on two nights during the first half of 2008, January 27 and April 24.

\section{Results}
\subsection{MAGIC}
\begin{figure}
 \centering
 \includegraphics[width=1.\linewidth]{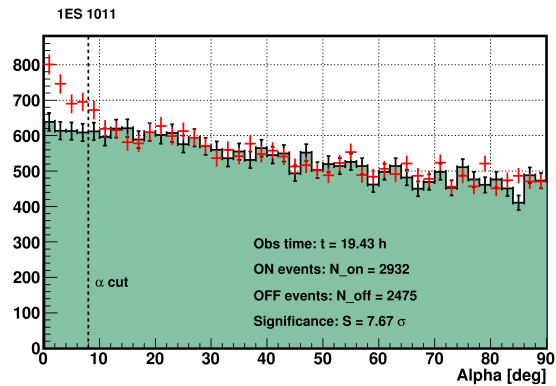}
 \caption{Distribution of $\alpha$ for ON-region and OFF-region data. ON-data are marked by the red crosses while the filled region are the OFF-data. The $\alpha$ cut is marked by the dashed line.}
 \label{Alpha_det}
\end{figure}

The MAGIC observations resulted in the confirmation of the source as a VHE emitter. The $\alpha$-plot shows 2932 ON-events in the $\alpha$ cut region of 8$^\circ$ (see Fig.~\ref{Alpha_det}). This and all other cuts applied in the VHE analysis were optimised on a sample of Crab Nebula data from the same epoch. A single OFF region directly opposite the ON region (with respect to the camera centre) was used to determine a background level of 2475 OFF-events applying the same event selection cuts. The calculated event excess corresponds to a statistical significance of 7.7\,$\sigma$ using equation (17) in ~\citet{lima1983}.

\begin{figure}
 \centering
 \includegraphics[height=1.\linewidth,angle=-90,keepaspectratio=true]{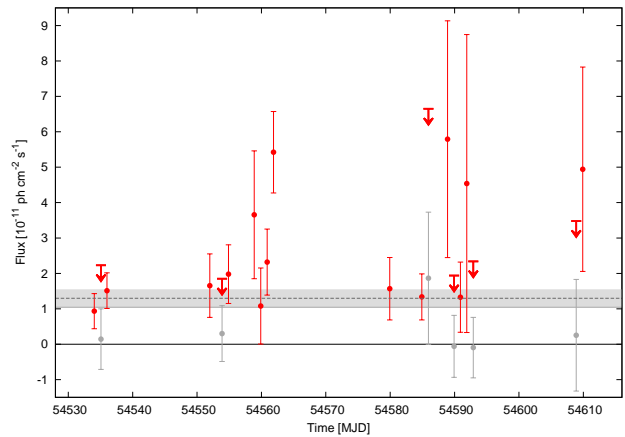}
 \caption{VHE light curve of 1ES\,1011$+$496. The daily integral fluxes above 200\,GeV are plotted as red filled ($>1$ sigma) and grey filled ($<1$ sigma) circles. The red arrows depict the 95 per cent confidence level upper limits computed for the observations yielding flux measurements with less than 1 sigma significance. The grey dashed line and the grey band represent the resulting fit to the data points with a constant flux along with its statistical uncertainty}.
 \label{vheLC}
\end{figure}

The daily VHE light curve of 1ES\,1011+496 showing the integral flux above 200 GeV of the source throughout the MAGIC observations can be seen in Fig.~\ref{vheLC}. In total, it consists of 476 excess events over 3293 background events. A fit with a constant to the data points yields $\chi^2$/d.o.f. $30.5/19$, corresponding to a probability of 19.5 per cent. The measured flux at MJD 54562 is $\sim3\sigma$ away from this fit, possibly indicating a higher flux on that night.
The mean flux above 200\,GeV is $(1.3\,\pm\,0.3)\,\times\,10^{-11}$ photons $\mathrm{cm}^{-2}\mathrm{s}^{-1}$.

\begin{figure}
 \centering
 \includegraphics[width=1.\linewidth]{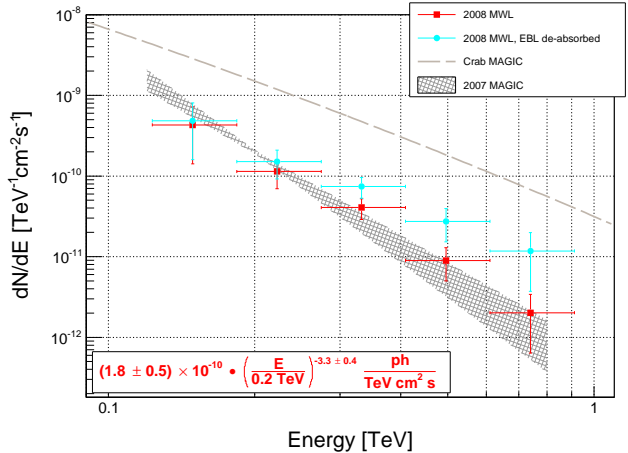}
 \caption{Measured spectrum of 1ES\,1011$+$496. The spectrum resulting from the 2008 observations is shown by the red squares, with the solid line representing a power-law fit to the data. The cyan data points show the deabsorped spectrum. The dark grey butterfly gives the spectrum measured during the discovery. For comparison the MAGIC Crab Nebula spectrum ~\citep{albert2008} is shown (the grey dashed line).}
 \label{spectrum}
\end{figure}

The energy threshold of the analysis was $\sim$100\,GeV enabling us to obtain a time-averaged VHE gamma-ray spectrum between $\sim$120 and $\sim$900\,GeV. The spectrum can be well described ($\chi^2\mathrm{/d.o.f.} = 0.7/3$ corresponding to a probability of 88 per cent) by a simple power law:
\begin{eqnarray*}
\frac{\mathrm {d}N}{\mathrm{d}E} = (1.8\pm 0.5)\left(\frac{E}{200\,\mathrm {GeV}}\right)^{-3.3\pm 0.4}\,\times10^{-10} \mathrm{TeV}^{-1}\,\mathrm{cm}^{-2}\,\mathrm{s}^{-1}.
\end{eqnarray*}
The resulting spectrum is shown in Fig.~\ref{spectrum}.

The derived mean VHE flux of $(1.8\pm 0.5)\times10^{-10} \,\mathrm{TeV}^{-1}\,\mathrm{cm}^{-2}\,\mathrm{s}^{-1}$ at 200 GeV is in a reasonably good agreement to the one published in the discovery paper by ~\citet{albert2007} who reported a flux of $(2.0\pm 0.1)\times10^{-10}\, \mathrm{TeV}^{-1}\,\mathrm{cm}^{-2}\,\mathrm{s}^{-1}$ at 200 GeV. Within the error bars also the spectral indices $\Gamma_{\mathrm{discovery}}=4.0\pm0.5$ and $\Gamma_{\mathrm{this work}}=3.3\pm0.4$ agree. The flux is approximately 10 times lower than reported from the source in February 2014 ~\citep{atel5887}.

Finally, we calculate the intrinsic VHE $\gamma$-ray spectrum, taking into account the absorption by $e^+e^{-}$ pair creation due to the interaction with the extragalactic background light (EBL) photons. Using the model of \citet{dominguez2011}, one of the several state-of-the-art EBL models, we derive an intrinsic spectral index of $\Gamma_{\mathrm int}=2.2\pm0.4$.

\begin{figure*}
 \centering
 \includegraphics[width=1.\linewidth]{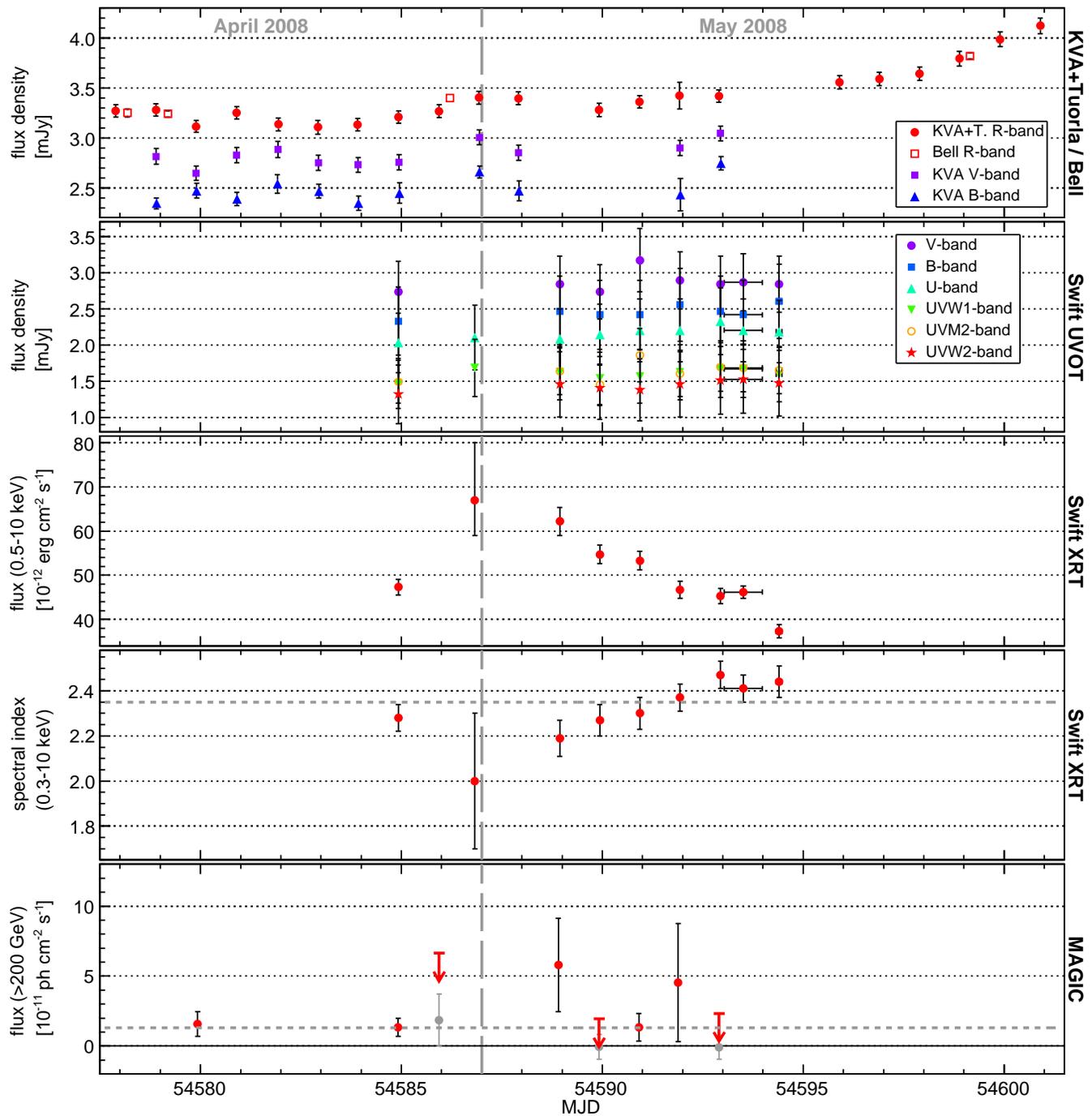}
 \caption{MWL light curve of 1ES\,1011$+$496 with daily binning. The optical and UV data (including the \textit{B}- and \textit{V}-band) from the KVA, Bell and UVOT telescopes are not host-galaxy corrected. In the \emph{Swift} spectral index and MAGIC plots the grey horizontal lines denote constant fits to the data points. In the bottom panel data shown in Fig. 3 are presented. The dashed grey vertical line denotes the transition between April and May 2008.}
 \label{1011_MWL}
\end{figure*}


\subsection{MeV-GeV}
AGILE did not detect the source during the campaign. The AGILE maximum likelihood analysis of the AGILE-GRID data taken during the first observation window yields a 95\% c.l. UL on the flux above 100\,MeV of $1.3 \times 10^{-7}\,\mathrm{photons}\,\mathrm{cm}^{-2}\,\mathrm{s}^{-1}$ with an effective exposure of about $9 \times 10^{7}\,\mathrm{cm}^2\,\mathrm{s}$. The effective exposure during the second observation window was too short to derive a meaningful UL. A search for source flares on time scales of 7 as well as 28 days using the entire AGILE-LV3 archive at ADC\footnote{The standard AGILE-LV3 archive is composed by counts, exposure and diffuse $\gamma$-ray background maps above 100 MeV generated on different timescales, and obtained from the official Level-2 data publicly available at the ADC site.}
up to the end of 2013, did not yield any detection with significance above 4$\sigma$.

1ES\,1011+496 is not very bright in MeV-GeV $\gamma$-rays, and in previous observations, EGRET did not clearly identify 1ES\,1011$+$496 during its entire mission~\citep{hartman1999}. The source is detected by AGILE above 4$\sigma$ significance level by integrating over a very long observation period of roughly 7 years, corresponding to an effective exposure of about 4.3 $\times 10^9$~cm$^2$~s. The estimated average $\gamma$-ray flux above 100 MeV is equal to $(5.4 \pm 1.4) \times$~10$^{-8}$~ph~cm$^{-2}$~s$^{-1}$, which is in agreement with the flux between 100 MeV and 100 GeV, as derived from the second Fermi-LAT Catalog, ~\citep[2FGL,][]{nolan2012}. In the \emph{Fermi}-LAT era the source has been included in the Bright AGN Sample ~\citep{abdo2009} as well as in the first ~\citep[1FGL,][]{abdo2010} and second ~\citep[2FGL,][]{nolan2012} \emph{Fermi}-LAT Catalog. Contrary to 1FGL, the source was characterized in 2FGL as significantly variable, which becomes also evident from comparing the spectral indices ($\Gamma_\mathrm{1FGL}=1.93\pm0.04$, $\Gamma_\mathrm{2FGL}=1.72\pm0.04$) and integral fluxes ($F_\mathrm{1-100 GeV, 1FGL}=\left(8.7\pm0.6\right) \times 10^{-9}\,\mathrm{ph}\,\mathrm{cm}^{-2}\,\mathrm{s}^{-1}$, $F_\mathrm{1-100 GeV, 2FGL}=\left(7.8\pm0.3\right) \times 10^{-9}\,\mathrm{ph}\,\mathrm{cm}^{-2}\,\mathrm{s}^{-1}$) derived from a simple power-law fit. It should be noted, however, that in 2FGL a log-parabolic power law is the preferred description of the 1ES\,1011$+$496 LAT spectrum, which cannot be attributed solely to absorption on the extragalactic background light ~\citep{ackermann2011}.

\subsection{X-rays}
At X-rays, the source has previously been detected by \emph{Einstein} ~\citep{elvis1992}, \emph{ROSAT} ~\citep{lamer1996, voges1999} and, more recently, by \emph{Swift} XRT~\citep{massaro2008, tavecchio2010, abdo2010, giommi2012}. Only ~\citet{lamer1996} and ~\citet{giommi2012} reported a steep power-law spectrum to be found at X-rays, the data of the remaining observations were better described by a broken power-law or log-parabolic fit. The reported peak energies vary between 0.04 and 0.74\,keV. 1ES\,1011$+$496 is characterized by strong variability at X-rays, with nearly a factor 20 difference reported for the integral flux between 2 and 10\,keV ($F_\mathrm{2-10 keV, low}=0.36 \times 10^{-11}\,\mathrm{erg}\,\mathrm{cm}^{-2}\,\mathrm{s}^{-1}$, $F_\mathrm{2-10 keV, high}=6.67 \times 10^{-11}\,\mathrm{erg}\,\mathrm{cm}^{-2}\,\mathrm{s}^{-1}$;~\citet{massaro2008, giommi2012}).

\begin{figure}
 \centering
 \includegraphics[width=0.7\linewidth,angle=270]{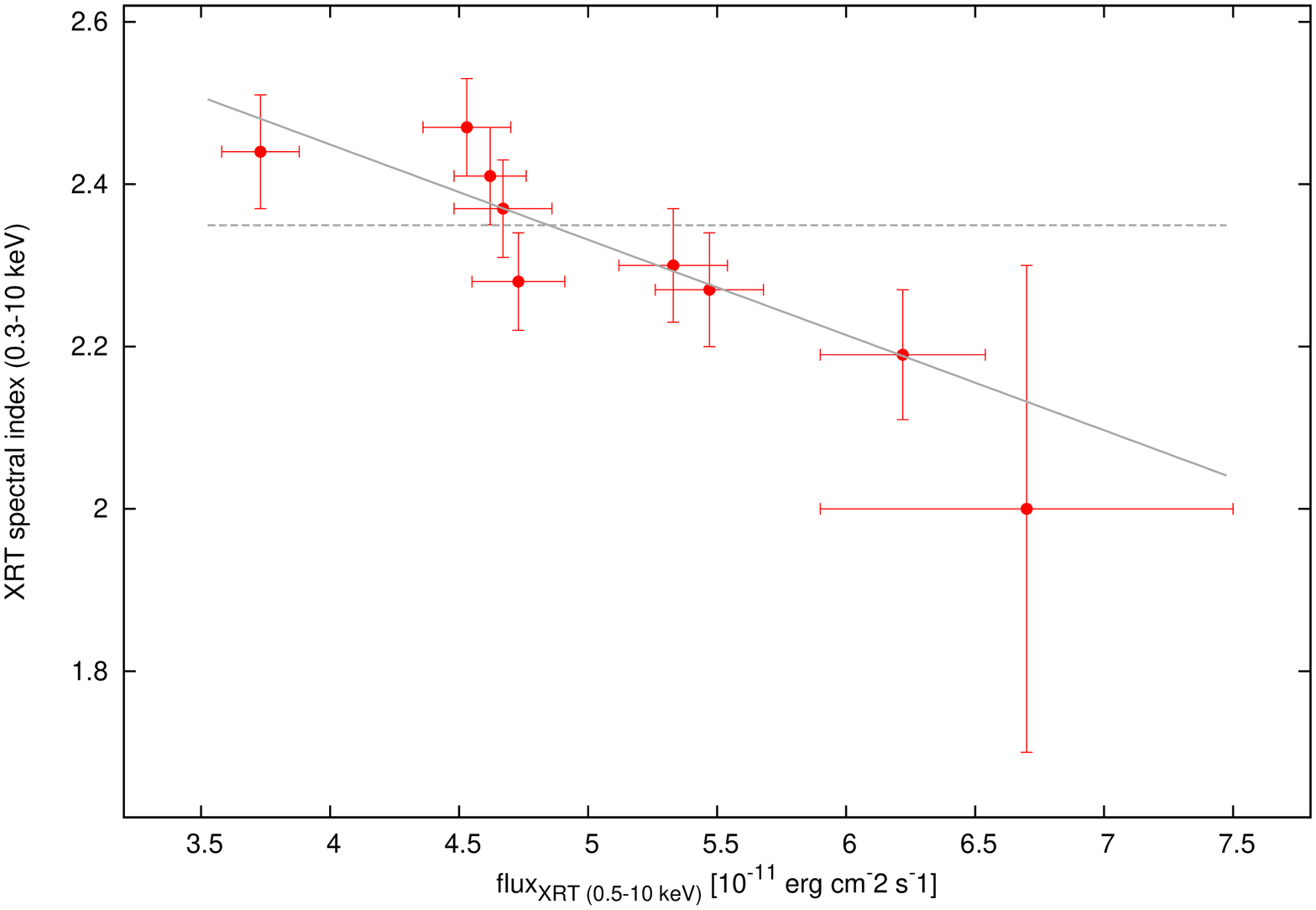}
 \caption{Simple power law spectral index determined between 0.3 and 10\,keV for the \emph{Swift} XRT data as a function of the integral flux between 0.5 and 10\,keV. A fit with a constant and a linear function is shown as the grey dashed and solid line, respectively.}
 \label{fig:XRT_flux_vs_idx}
\end{figure}

During this campaign an X-ray flare was clearly detected, as can be seen in the light curve in Fig.~\ref{1011_MWL}. The observation sampling prevented to define a baseline flux level, therefore the rise and fall times of the flare could not be evaluated from these measurements alone. Also the peak of the flare cannot be defined accurately, considering that the highest flux point is consistent within the error bars with the second-largest measurement. Compared to archival observations, the source was found in rather high flux states ranging from $F_\mathrm{0.5-10 keV} = 3.73 \times 10^{-11}\,\mathrm{erg}\,\mathrm{cm}^{-2}\,\mathrm{s}^{-1}$ to $F_\mathrm{0.5-10 keV} = 6.70 \times 10^{-11}\,\mathrm{erg}\,\mathrm{cm}^{-2}\,\mathrm{s}^{-1}$.
The spectral index was not significantly variable during the flare, a fit with a constant yielding a $\chi^2$/d.o.f.\ of 15.3/8 (see Fig.~\ref{1011_MWL}). However, the integral flux variability seems to be correlated with the spectral index(linear fit is favoured over the constant one with 98.8 per cent likelihood), as shown in Fig.~\ref{fig:XRT_flux_vs_idx}. Such a harder-when-brighter trend has often been found for BL Lac objects (e.g.~\citet{giommi1990, pian1998, acciari2011}), but is reported here for the first time for 1ES\,1011$+$496. 

\emph{Swift} UVOT observed the source on all filters (\textit{V, B, U, UVW1, UVM2, UVW2}). The two shortest exposures were insufficient to derive a flux for all but the \textit{U}- and \textit{UVW1}-band in case of the second shortest exposure. The \textit{V}- and \textit{B}-band results were well compatible with the contemporaneous KVA data (see below). However, significant variability could not be detected by UVOT in any band during the campaign (see Fig.~\ref{1011_MWL}), which can be ascribed to the rather large uncertainty of the measurements.

\subsection{Optical}
The object shows variability in its optical brightness with a core flux increase detected by the Tuorla Blazar Monitoring Program by up to a factor of $\sim$2--3 over the lowest states during flares. During the campaign the source was in a relatively high state in the optical \textit{R}-band with a mean flux of $(3.33\pm0.06)$ mJy. This is $\sim$20 per cent lower than the $(4.14\pm0.04)$ mJy average measured in 2007 during the MAGIC discovery, which was triggered by an optical flare \citep{albert2007}. Throughout the campaign, the optical flux displayed on average an increasing trend and crossed the flare threshold calculated in ~\citet{reinthal2012a} around half-way into the campaign. From the end of April to the beginning of May the source was also observed in the \textit{V}- and \textit{B}-bands. The fluxes in these bands followed in general the same trends as in the \textit{R}-band. 
The optical observations by Bell were compatible with the KVA data, showing the same trend in flux density.

We also constructed optical SEDs using the KVA and UVOT data that were not separated by more than 1 hour from each other. The data were host galaxy subtracted and de-reddened (see Sections 2.3 and 2.4). The resulting SEDs are discussed in detail in Section 4.3.

The three polarization measurements taken on MJD 54583, 54584 and 54593 all show a low polarization ($2.6\pm0.9, 2.2\pm0.8$ and $2.5\pm0.4$ per cent) and a rather stable electric vector position angle ($139\pm10, 165\pm10, 153\pm4$ degrees). The low polarization is in agreement with previous observation ~\citep{wills2011}. 

\subsection{Radio}
Previous observations of the source have reported variability at the radio bands, with flux densities at 1.4\,GHz varying between $\sim$380\,mJy and $\sim$470\,mJy~(\citet{nakagawa2005} and references therein). At 37\,GHz, the Mets\"ahovi radio telescope detected the source only once from 12 pointings between 2002 and 2005, measuring a flux density of $\left(0.56\pm0.12\right)$\,Jy on December 1, 2002 ~\citep{nieppola2007}. No detection was achieved five days later, which may either be a sign of rather fast variability, or a consequence of the observation condition-dependent detection limit of Mets\"ahovi. Around this campaign 1ES\,1011$+$496 was not detected by Mets\"ahovi. For the second observation on 2008 April 24 (MJD 54580), an UL on the flux density at 37\,GHz of $<$\,0.62\,Jy (S/N $>$ 4) was calculated. This value is compatible with the detections achieved at the end of 2008 ($F_\mathrm{37 GHz} \approx \left(0.45 \pm 0.10\right)$\,Jy), where a significant signal was observed on three occasions within five days. This is the kind of behaviour that both IBL and HBL sources typically show at 37 GHz \citep{nieppola2007}. They seem to spend most of the time below the detection limit, and are only occasionally detected.

We also investigated archival and publicly available VLBI data of the source. 1ES\,1011$+$496 exhibits a core-jet morphology typical for blazars, with no sign of a counter-jet and a jet position angle well-aligned on pc and kpc scales~\citep{augusto1998, nakagawa2005}. From two VLBI observations spanning 2.2\,years, no obvious jet motion was visible ~\citep{nakagawa2005}. ~\citet{kharb2008} report a fractional core polarization of $\geq 4$ per cent, compared to $<$\,3 per cent for other HBLs studied. This rather high value is confirmed by studying public MOJAVE\footnote{\href{http://www.physics.purdue.edu/MOJAVE/}{http://www.physics.purdue.edu/MOJAVE/}} polarization observations of the core$+$jet, ranging from 2.9\% to 8.1\%. 


\begin{figure*} 
 \centering
 \includegraphics[width=.57\linewidth, angle=270]{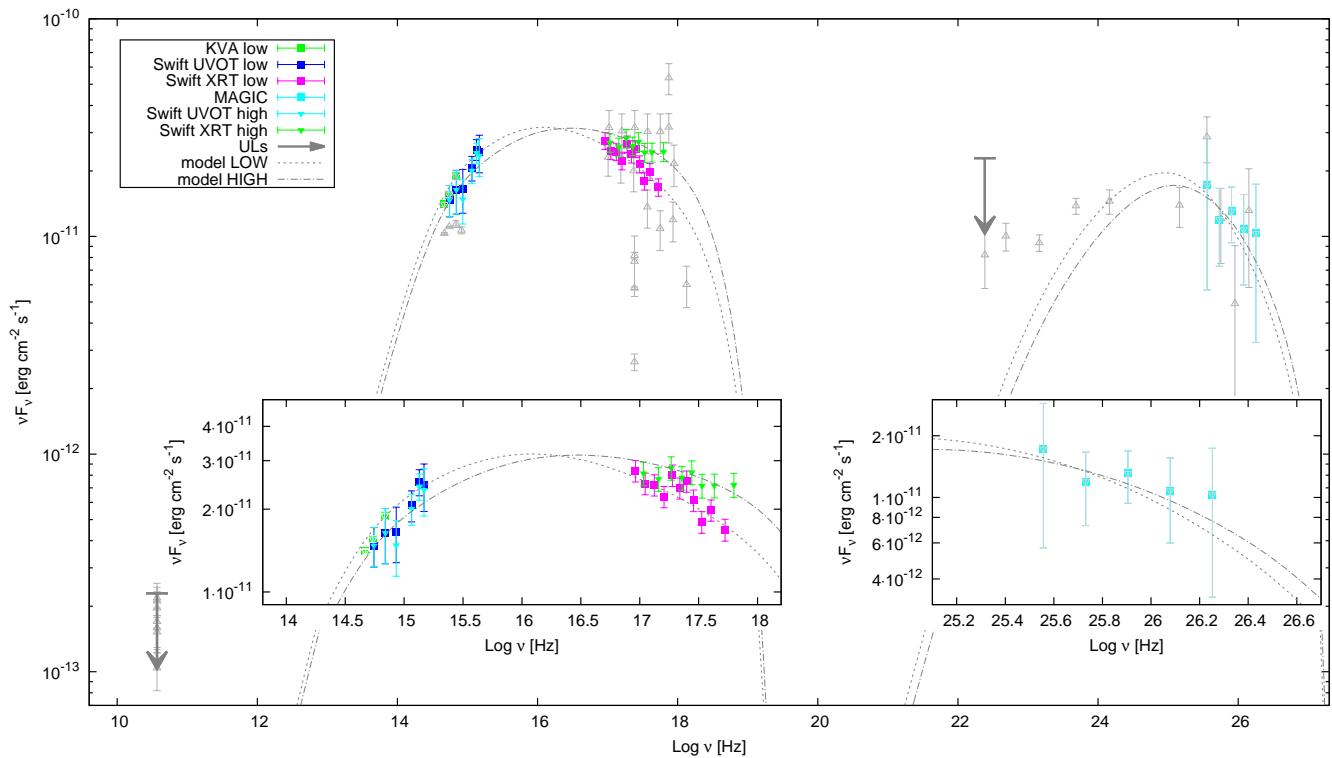}
 \caption{MWL SEDs of 1ES\,1011$+$496. The triangles and squares depict the high and low X-ray states of MJD 54589 and MJD 54593, respectively. The KVA low and UVOT data are corrected for Galactic absorption~\citep{schlafly2011} and the optical bands also for host-galaxy contribution~\citep{nilsson2007}. The Mets\"ahovi 37\,GHz and AGILE $\gamma$-ray upper limits from this campaign are shown with arrows.
The \emph{Fermi}-LAT data from the LAT 1-year Catalog ~\citep{abdo2010} along with the other non-simultaneous data are added for illustrative purposes (grey symbols). The MAGIC data have been corrected for EBL absorption using the model of \citet{dominguez2011}. See main text for a description of the model curves.}
 \label{1011_SED}
\end{figure*}

\begin{table*}
 \centering
 \caption{\small{One-zone SSC model parameters for high X-ray and low X-ray states}} 
  \begin{tabular}{c|ccccccccc}
  \hline
  \hline
  Model & $B$ & $\delta$ & $R$ & $K$ & $p_1$ & $p_2$ & $\gamma_\mathrm{min}$ & $\gamma_\mathrm{break}$ & $\gamma_\mathrm{max}$ \\
  & [G] & & [$10^{16}$ cm] & [cm$^{-3}$] & & &[$10^3$] & [$10^4$] & [$10^5$] \\
  \hline
 high X-ray & 0.048 & 26 & 3.25 & 700 & 1.9 & 3.3 & 7.0 & 3.4 & 8.0 \\
 low  X-ray & 0.048 & 26 & 3.25 & 800 & 1.9 & 3.5  & 7.0 & 3.4 & 8.0 \\ 
  \hline
  \end{tabular}
 \label{table_param}
\end{table*}

\section{Discussion}
\subsection{Multi-wavelength light curve}
Looking at the MWL light curve we see moderate variability in the optical and X-ray bands with the observations in the latter capturing part of the rise and decay phases of an X-ray flare. The flux in VHE $\gamma$-rays is consistent with being constant.

The investigation of the simultaneous light curves for correlations between different energy bands is hindered by the rather sparse sampling and non-significant variability in the VHE $\gamma$-ray band. The few simultaneous data pairs are not sufficient to establish a meaningful connection between the optical and X-ray, X-ray and VHE or optical and VHE light curves (see Fig.~\ref{1011_MWL}). Serendipitous observations by \emph{INTEGRAL} ISGRI between MJD 54589 and MJD 54593 did not yield a significant detection of the source, which seems to be the case also for \emph{Swift}-BAT and \emph{RXTE}-ASM\footnote{judging from the publicly available quick-look light curves (\href{http://xte.mit.edu/ASM\_lc.html}{http://xte.mit.edu/ASM\_lc.html}, \href{http://swift.gsfc.nasa.gov/docs/swift/results/transients/}{http://swift.gsfc.nasa.gov/docs/swift/results/transients/})}. However, the good coverage in energy allows to investigate the contemporaneous broad-band SED of 1ES\,1011$+$496.

\subsection{Spectral energy distribution}
Two separate data sets, 'high X-ray' and 'low X-ray', defined according to the X-ray flux state of the source on MJD  54588.9 (high) and MJD 54592.9 (low) as can be seen in Fig.~\ref{1011_MWL} and availability of quasi-simultaneous ($\pm0.5$ days) MWL data, were used to construct quasi-simultaneous SEDs (see Fig.~\ref{1011_SED}). 
The VHE results do not show significant variability in the course of the campaign. Therefore, to reduce the error bars of the measurement, the time-averaged spectrum is used for the SED. Due to the small second AGILE observation window, no AGILE upper limit could be extracted for the periods from which the SEDs were calculated and the upper limit from first period is shown. {\it Fermi}-LAT was not yet operating at the time of the campaign. Although the  {\it Fermi}-LAT found 1ES\,1011$+$496 to be significantly variable after 24 month of observations ~\citep{nolan2012}, we included the LAT 1FGL spectrum for SED modelling as an order of magnitude estimate of the flux in this energy regime. 

From the SED a basically equal power emitted by both the synchrotron and the SSC components can be seen. Even though the \emph{Fermi}-LAT data were not measured simultaneously to the other data, they connect well to the VHE spectrum which is comparable to the discovery spectrum.  
At lower energies, the optical flux measured by KVA was found to be lower by $\sim$20 per cent, and 
the X-ray flux was almost 10 times higher than the archival measurements used
in the VHE discovery paper. 
The low (non-simultaneous) X-ray flux constraining the SED modelling led ~\citet{albert2007} to the conclusion that in this source, the inverse-Compton component would dominate over the synchrotron component. On the contrary, the quasi-simultaneous SEDs from our MWL observations indicate that this interpretation may not be correct, corroborating that this source is synchrotron dominated like most HBLs.

The data were modelled using a one-zone SSC model ~\citep{maraschi2003}. 
It assumes a relativistically moving emission region characterised by its radius $R$, magnetic field $B$ and Doppler factor $\delta$. The emission region contains an electron population with normalization $K$ at $\gamma$ = 1 following a broken power-law distribution with index $p_1$ for $\gamma_\mathrm{min}<\gamma<\gamma_\mathrm{break}$ and $p_2$ for $\gamma_\mathrm{break}<\gamma<\gamma_\mathrm{max}$. This one-zone model cannot reproduce the spectrum at the lowest frequencies, since the emission is self-absorbed below the millimeter band. It is generally assumed that this part of the SED is due to outer regions of the jet. 

Fig.~\ref{1011_SED} shows the results of the SSC modelling describing the quasi-simultaneous high X-ray and low X-ray state SEDs. The SSC model parameters are reported in Table 3. The goodness of the model is judged by eye and hence the curve represents only one possible set of SED parameters, instead of being a real fit to the data. The small difference between the high X-ray and low X-ray state SEDs, 
is modelled as a slight steepening of the high energy electron spectrum and a slight increase in the electron number density, which can be interpreted as cooling of the emitting electrons with a small injection, mostly related to the lowest energies considered here.

The model curves describe the optical, X-ray and VHE $\gamma$-ray data rather well. 
To reproduce the narrowness of the synchrotron peak in the low X-ray state a narrow electron energy distribution with large $\gamma_\mathrm{min}$ and small $p_1$ is required. Such narrow synchrotron peaks have been found also previously, e.g. in 1ES\,1215$+$303 \citep{aleksic2012}.

The non-simultaneous {\it Fermi}-LAT data are not well described by the model neither in the low nor in the high X-ray state. The {\it Fermi}-LAT data would require the inverse-Compton peak to be broader than the synchrotron peak which is difficult to model within the adopted one-zone model. Especially at low $\gamma$-ray energies a discrepancy with the model becomes evident. However, that discrepancy is partly alleviated considering that the 1FGL is integrated over several months and the shape is a sum of several flux and spectral states. Additionally, a second, more Compton dominated component may contribute to the discrepancy. To investigate this, simultaneous {\it Fermi} and MAGIC observations are needed and were performed in 2011/2012 \citep{aleksic16}.

\citet{2015A&A...573A...7W} have used preliminary results of this campaign to test their self-consistent and time-dependent hybrid blazar emission model. Requiring $p_2 - p_1\sim$1 for the cooling break they concluded that their one-zone SSC model cannot reproduce the narrow shape of the synchrotron peak as long as the magnetic field is weak. However, we note that the condition $p_2 - p_1\sim1$ only holds in the case of shock acceleration and synchrotron and/or inverse Compton (in Thomson regime) cooling on a perfectly uniform and homogeneous region. Furthermore, the break is not necessarily related to cooling, but can be intrinsic to the electron energy distributions, probably caused by a decrease in the efficiency to accelerate the electrons at the highest energies. With the increase in the simultaneity and quality of the measured SEDs (covering now a larger portion of the electromagnetic spectrum), a number of recent works showed the requirement of a change of index larger than the canonical value of 1 for a large number of sources \citep[e.g.][]{abramowski2013,2014A&A...563A..90A,2014A&A...567A.135A,2014A&A...572A.121A,2015MNRAS.450.4399A,2015A&A...578A..22A,2015MNRAS.451..739A}.
Therefore, one-zone SSC models cannot be excluded in general, and with the set of parameters shown in Table~\ref{table_param} such a model reproduces well the shape of the SED of 1ES 1011+496 during high and low activity. Compared to values derived for a large sample of TeV detected BL Lac objects using a leptonic one-zone SSC model ~\citep{tavecchio2010}, most of the parameters do not deviate significantly. 

\subsection{Source classification}

\begin{figure}
 \centering
 \includegraphics[width=.7\linewidth, angle=270]{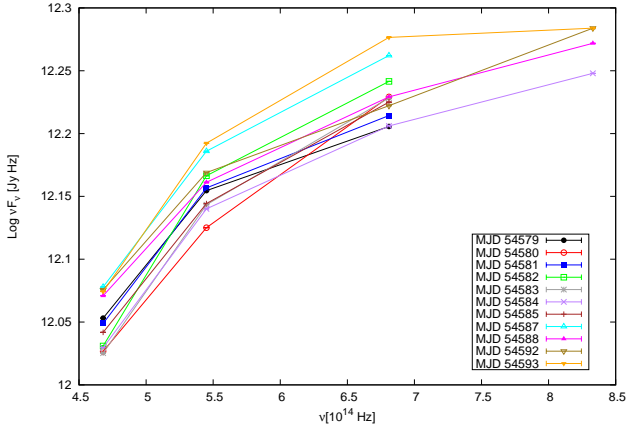}
 \caption{Daily optical spectral energy distributions of 1ES\,1011$+$496 from the optical \textit{U}-, \textit{B}-, \textit{V}- and \textit{R}-bands.}
 \label{Opt_sed}
\end{figure}

Although historically classified as an HBL \citep{donato2001, nieppola2006, abdo2010} and showing often hard X-ray spectra, the report by~\citet{bottcher2010} of synchrotron peak frequencies in the optical range suggests an IBL nature of 1ES\,1011$+$496. Moreover, the deviation from a simple power-law behaviour reported in 2FGL\footnote{Out of 69 sources best described by a curved spectrum in the LAT range, 1ES\,1011$+$496 is the only HBL ~\citep{ackermann2011}.}, the at times steep X-ray spectra, the presence of a superluminal jet component ~\citep{lister2013} and the rather high core polarization seen in the radio~\citep{kharb2008} point to an IBL rather than an HBL object. This ambiguity could be explained considering that 1ES\,1011$+$496 shows a trend of a higher peak frequency with increasing flux in the optical as well as X-ray regime. Hence it would be natural to assume that the basic classification of the source is in fact IBL, as observed at optical frequencies during low to medium flux states and which seems to dominate the (time-averaged) 2FGL HE spectrum, whereas during high flux states, the synchrotron peak shifts to frequencies characteristic of HBLs. It is known that in so-called extreme blazars, the synchrotron peak may shift by more than an order of magnitude (e.g.~\citet{pian1998, giommi2000, costamante2001}) during flares and therefore it is expected that such objects would exist. To our knowledge the explanation of these at first glance contradictory phenomena as an underlying IBL nature of the object showing HBL features during high states has not been made before in the literature. However, it should be noted that 1ES~1011+496 was not the first source for which such features have been reported, since \citet{abramowski2013} observed in PKS~0301--243 that the first peak is located at a very low frequency (assuming an HBL nature) and that the position is shifting with increasing flux above the formal boundaries of IBLs towards HBLs.

\begin{figure}
 \centering
 \includegraphics[width=.7\linewidth, angle=270]{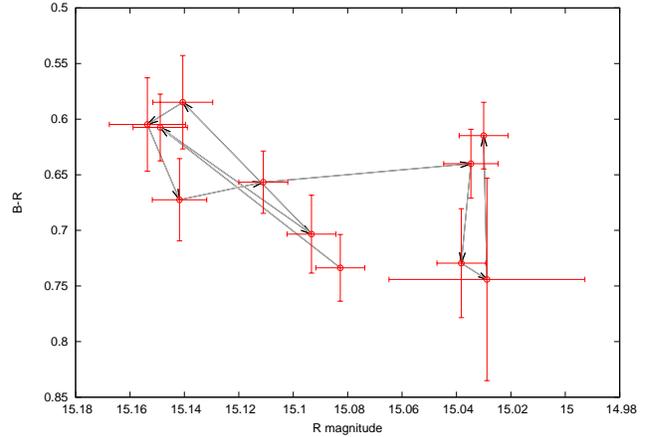}
 \caption{Colour-magnitude diagram for 1ES\,1011$+$496 from MJD 54579 to MJD 54593. The grey line with arrows traces the evolution of colour with time.}
 \label{B-R_mag}
\end{figure}

In order to shed light into the nature of the source, we constructed optical SEDs using the KVA and UVOT data that were not separated by more than 1 hour from each other. The resulting 11 SEDs are shown in Fig.~\ref{Opt_sed}. During all epochs the optical SED of 1ES\,1011$+$496 shows an increasing trend, suggesting that the synchrotron peak is located at a frequency above $10^{15}$ Hz. This is in agreement with the HBL classification and in contradiction with the results of~\citet{bottcher2010}, although the observations in that paper are partially from the same period. We suggest that the contradiction might, at least partially, originate from host galaxy subtraction, which was neglected in~\citet{bottcher2010}.

We also studied the dependence of the \textit{B-R} color index on the brightness of the source using the same criterion for selecting the data pairs as described above. The resulting color-magnitude diagram plotting the \textit{B-R} color as a function of \textit{R}-band magnitude is shown in Fig.~\ref{B-R_mag}. Unlike in~\citet{bottcher2010} who noted a bluer-when-brighter trend, no dependence on the B-R color on the source brightness was found. A fit with a constant to the data yields a $\chi^{2}\mathrm{/d.o.f.}$ of 21.6/10 while a linear fit gives a $\chi^{2}\mathrm{/d.o.f.}$ of 21.1/9, corresponding to probabilities of $\sim1$--2 per cent for both fits. We also searched for an evolution in time in the colour-magnitude diagram, but within the error bars no pattern was found.

We conclude that in our optical study the source behaves like a typical HBL rather than IBL. It should be noted, however, that both the color-magnitude diagram and the optical SEDs calculated in this paper cover a rather narrow brightness range (time period between MJD 54578 to 54595, for which we have multiband optical-UV observations) and are representative of a rather high state of the source. Therefore in order to conclusively determine the true classification of the source it should be studied with further multi-band optical observations covering a larger range of flux states than presented here. 

As reported in \cite{atel5887}, the VHE activity of 1ES\,1011+496 increased by one order of magnitude in 2014, which triggered further observations at different wavelengths. A short study using the measured VHE spectra to constrain the extragalactic background density has recently been published \cite{Ahnen2016}, but the results from the multi-instrument observations are not yet available. The study of the broadband SED of 1ES\,1011+496 during this very high activity period in 2014, together with a detailed comparison with the broadband SED reported in this paper will help to understand better this very interesting blazar.

\section{Summary and Conclusion}
In this paper we report the first MWL campaign including VHE coverage on the blazar 1ES\,1011$+$496. The campaign was performed 
regardless of the 
state of the source. Compared to archival data the source appeared to be in a rather high state during these observations.

The MAGIC observations presented in this paper confirm the source as a VHE $\gamma$-ray emitter.
The VHE flux was found to be at a similar level to that measured during the discovery and consistent with being constant. The HE $\gamma$-ray observations with AGILE did not yield a detection of the source, with flux upper limits in agreement with the average flux state observed by {\it Fermi}-LAT.

In X-rays the source was variable during the campaign with the observations catching part of the rise and decay phases of a flare and showed a harder-when-brighter behaviour as often seen for sources of this type. The X-ray emission was also slightly higher and harder than that of the archival data. 

In the optical band the source was in a slightly lower state (by $\sim$20 per cent) than during the VHE discovery that was triggered by an optical high state. However, a potential optical--VHE connection cannot be assessed from these observations, since the 20 per cent difference detected at optical is well within the statistical uncertainties of these MAGIC observations.

We performed a detailed optical study of 1ES\,1011$+$496 in order to determine the IBL/HBL nature of the source suggested by a study of archival data. The host galaxy subtracted SEDs show a clear increasing trend indicating that the synchrotron peak is not located in the optical band and no magnitude-dependence of the \textit{B-R} colour index was found, both contradicting the findings of~\citet{bottcher2010} and the suggested IBL nature. However, the results presented here are derived from a relatively narrow range of flux states and further multi-band optical observations extending to both higher and lower source states are necessary to answer this question. Despite that, it is clear that VLBI radio data and HE $\gamma$-ray data (see 4.3.) point towards a behaviour untypical for HBLs and the source seems to be a borderline case between IBL and HBL. It has been suggested that IBLs and HBLs are intrinsically the same objects with similar jet powers, the difference originating from a larger misalignment of the IBL jets to our line of sight ~\citep{2011ApJ...740...98M} and therefore it would not be surprising if sources in the borderline existed.

We also constructed the first quasi-simultaneous broad band spectral energy distribution of the source with VHE coverage. These observations show that, unlike concluded in \citet{albert2007}, the synchrotron and IC peaks have similar powers, i.e.\,the source is not Compton-dominated and therefore a typical VHE $\gamma$-ray emitting BL Lac object. A one-zone SSC model describes the observed contemporaneous SED relatively well, yielding rather typical values for a VHE BL Lac object despite the narrowness of the synchrotron peak requiring a rather narrow electron energy distribution, which restricted the SED model parameter space.  
However, we note that the spectral energy distribution modelling will highly benefit from simultaneous HE to VHE $\gamma$-ray observations, which were conducted in 2011 and 2012 
 as a follow-up of this campaign \citep{aleksic16}.\\

\section*{Acknowledgements}
We would like to thank
the Instituto de Astrof\'{\i}sica de Canarias
for the excellent working conditions
at the Observatorio del Roque de los Muchachos in La Palma.
The financial support of the German BMBF and MPG,
the Italian INFN and INAF,
the Swiss National Fund SNF,
the ERDF under the Spanish MINECO (FPA2012-39502), and
the Japanese JSPS and MEXT
is gratefully acknowledged.
This work was also supported
by the Centro de Excelencia Severo Ochoa SEV-2012-0234, CPAN CSD2007-00042, and MultiDark CSD2009-00064 projects of the Spanish Consolider-Ingenio 2010 programme,
by grant 268740 of the Academy of Finland,
by the Croatian Science Foundation (HrZZ) Project 09/176 and the University of Rijeka Project 13.12.1.3.02,
by the DFG Collaborative Research Centers SFB823/C4 and SFB876/C3,
and by the Polish MNiSzW grant 745/N-HESS-MAGIC/2010/0. \\
The AGILE Mission is funded by the Italian Space Agency (ASI), with scientific and programmatic participation by the Italian Institute of Astrophysics (INAF) and the Italian Institute of Nuclear Physics (INFN). Research partially supported through the ASI grants no. I/089/06/2, I/042/10/0 and I/028/12/0. \\
We gratefully acknowledge the entire \emph{Swift} team, the duty scientists and science planners for the dedicated support, making these observations possible.\\
This research has made use of the XRT Data Analysis Software (XRTDAS) developed under the responsibility of the ASI Science Data Center (ASDC), Italy. \\
M.~T.~Carini, Richard Walters, April Pease acknowledge support from the Institute for Astrophysics and Space Science at Western Kentucky University.\\
The Mets\"ahovi team acknowledges the support from the Academy of Finland to our observing projects (numbers 212656, 210338, 121148, and others).\\ 
This research has made use of NASA's Astrophysics Data System.\\
Part of this work is based on archival data, software or on-line services provided by the ASI Science Data Center (ASDC).\\ 
This research has made use of the NASA/IPAC Extragalactic Database (NED) which is operated by the Jet Propulsion Laboratory, California Institute of Technology, under contract with the National Aeronautics and Space Administration.\\ 
This research has made use of the SIMBAD database, operated at CDS, Strasbourg, France.\\
We thank the anonymous referee for a careful reading and useful
questions and suggestions that improved the presentation of this paper.\\

\bibliographystyle{mn2e}
\bibliography{bibfile}

\vspace{0.5cm}
\begin{minipage}[t]{0.5\textwidth} 
$^{1}$ {ETH Zurich, CH-8093 Zurich, Switzerland} \\
$^{2}$ {Universit\`a di Udine, and INFN Trieste, I-33100 Udine, Italy} \\
$^{3}$ {INAF National Institute for Astrophysics, I-00136 Rome, Italy} \\
$^{4}$ {Universit\`a  di Siena, and INFN Pisa, I-53100 Siena, Italy} \\
$^{5}$ {Croatian MAGIC Consortium, Rudjer Boskovic Institute, University of Rijeka and University of Split, HR-10000 Zagreb, Croatia} \\
$^{6}$ {Saha Institute of Nuclear Physics, 1\textbackslash{}AF Bidhannagar, Salt Lake, Sector-1, Kolkata 700064, India} \\
$^{7}$ {Max-Planck-Institut f\"ur Physik, D-80805 M\"unchen, Germany} \\
$^{8}$ {Universidad Complutense, E-28040 Madrid, Spain} \\
$^{9}$ {Inst. de Astrof\'isica de Canarias, E-38200 La Laguna, Tenerife, Spain; Universidad de La Laguna, Dpto. Astrof\'isica, E-38206 La Laguna, Tenerife, Spain} \\
$^{10}$ {University of \L\'od\'z, PL-90236 Lodz, Poland} \\
$^{11}$ {Deutsches Elektronen-Synchrotron (DESY), D-15738 Zeuthen, Germany} \\
$^{12}$ {IFAE, Campus UAB, E-08193 Bellaterra, Spain} \\
$^{13}$ {Universit\"at W\"urzburg, D-97074 W\"urzburg, Germany} \\
$^{14}$ {Centro de Investigaciones Energ\'eticas, Medioambientales y Tecnol\'ogicas, E-28040 Madrid, Spain} \\
$^{15}$ {Universit\`a di Padova and INFN, I-35131 Padova, Italy} \\
$^{16}$ {Institute for Space Sciences (CSIC\textbackslash{}IEEC), E-08193 Barcelona, Spain} \\
$^{17}$ {Technische Universit\"at Dortmund, D-44221 Dortmund, Germany} \\
$^{18}$ {Unitat de F\'isica de les Radiacions, Departament de F\'isica, and CERES-IEEC, Universitat Aut\`onoma de Barcelona, E-08193 Bellaterra, Spain} \\
$^{19}$ {Universitat de Barcelona, ICC, IEEC-UB, E-08028 Barcelona, Spain} \\
$^{20}$ {Japanese MAGIC Consortium, ICRR, The University of Tokyo, Department of Physics and Hakubi Center, Kyoto University, Tokai University, The University of Tokushima, KEK, Japan} \\
$^{21}$ {Finnish MAGIC Consortium, Tuorla Observatory, University of Turku and Department of Physics, University of Oulu, Finland} \\
$^{22}$ {Inst. for Nucl. Research and Nucl. Energy, BG-1784 Sofia, Bulgaria} \\
$^{23}$ {Universit\`a di Pisa, and INFN Pisa, I-56126 Pisa, Italy} \\
$^{24}$ {ICREA and Institute for Space Sciences (CSIC\textbackslash{}IEEC), E-08193 Barcelona, Spain} \\
$^{25}$ {Universit\`a dell'Insubria and INFN Milano Bicocca, Como, I-22100 Como, Italy} \\
$^{26}$Universit\`a di Trieste, and INFN Trieste, I-34127 Trieste, Italy\\
$^{27}$ASI Science Data Center, I-00133 Rome, and INAF-Oar I-00040 Monteporzio Catone, Italy\\
$^{28}$INAF, Istituto di Astrofisica Spaziale e Fisica Cosmica, I-90146 Palermo, Italy\\
$^{29}$Tuorla Observatory, Department of Physics and Astronomy, University of Turku, V\"ais\"al\"antie 20, FI-21500 Piikki\"o, Finland\\
$^{30}$Department of Physics and Astronomy, Western Kentucky University, Bowling Green, KY 42103, USA\\
$^{31}$Aalto University Mets\"ahovi Radio Observatory, Mets\"ahovintie 114, FI-02540 Kylm\"al\"a, Finland\\
$^{32}$Aalto University Department of Radio Science and Engineering,  
P.O. BOX 13000, FI-00076 AALTO, Finland\\
$^{*}$ Corresponding authors: R. Reinthal, e-mail: rirein@utu.fi, S. R\"ügamer, e-mail: snruegam@astro.uni-wuerzburg.de, E. Lindfors, e-mail: elilin@utu.fi
\end{minipage}

\label{lastpage}

\end{document}